# Holophotography with a diffraction grating

José J. Lunazzi

Universidade Estadual de Campinas, Institute of Physics

C.P.6165

13083-950 Campinas SP, Brazil

Abstract

Color encoding of depth is shown to occur naturally in images of objects observed through diffraction gratings under common white light illumination. A synthetic image is then obtained from a single point of view, a phenomenon that can be applied to stereophotography. The image can be recorded in a common color photograph, providing a simple method of visual decoding by means of ordinary colored 3D spectacles. The fundamental equation and the photographic procedure for maximum fidelity in three-dimensional reproduction are described. The result is a photograph that has the capability of registering all the views of an object in a continuous sequence, which is called holophotography and was previously obtained by means of a hologram. By eliminating the need for a laser and holographic film, a new technique for holography in white light is foreseen.

*Subject terms: holography; photograph; stereoscopy; color encoding; three-dimensional imaging*

CONTENTS

## 1. INTRODUCTION

In a previous paper[1], it was shown that a continuous sequence of views of an object can be synthesized naturally by means of holography. Every view within a restricted angular range can be encoded by means of a characteristic wavelength value and be deflected by the hologram and sent to a single observation point.

Since the main mechanism for this phenomenon is diffraction, it is possible, as shown in the present paper, to produce an equivalent image by means of a diffraction grating, thus eliminating the need for a laser and holographic film or processing. Thus, we can use common color film to obtain a photograph with stereo properties that can be observed in 3-D by means of colored glasses. We can also perform animated reconstruction, as described previously for the photograph of a hologram of the object[1]. Briefly, the method is as follows: A synthetic image of a given object, comprising all possible images obtained by use of a diffraction grating and observable from a single point of view, is recorded on color film.

As observed visually, the main characteristic of the synthetic image is a spectral multicolor profiling of shapes in the scene, a blurring that resembles a severe degree of chromatic aberration. This blurring is more apparent for regions of the object that are more distant from the plane where the chromatic dispersion is generated, constituting a way of using color for encoding depth. There are various ways of extracting the desired colors of images for stereoscopic viewing. We present here the experimental setup, the fundamental equations, and the procedure for obtaining maximum fidelity in 3-D reproduction.

## 2. IMAGES OF A POINT OBJECT OBSERVED THROUGH A DIFFRACTION GRATING

Figure 1 shows a point object A, radiating white light to a transmission diffraction grating through which it is being observed. The coordinates of A are $x_A$, $z_A$, and the observation coordinates are $x_F$, $z_F$. Coordinate x indicates the incidence point at the grating.

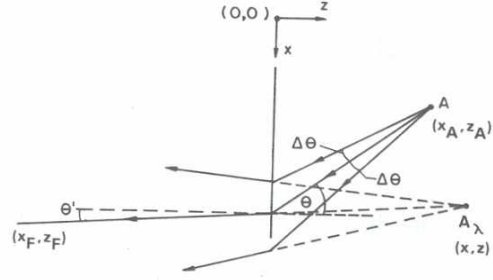

*Fig. 1. Ray tracing scheme for obtaining a diffracted image. A is the point object; the diffraction grating is at z=0.*

The basic equation for diffraction of a light ray generated from point A impinging on the grating at angle θ and emerging from the grating at angle θ' is

$$\sin\theta - \sin\theta' = \frac{j\lambda}{d} \qquad (1)$$

where d is the period of the grating, j is the diffraction order number, and A is the wavelength.

The equation of a line having the same direction as the diffracted ray of light and coincident with it can be expressed as

$$z = \frac{z_A \tan(\theta) - x + x_A}{T(\theta)} \qquad (2)$$

where

$$T(\theta) = \tan(\arcsin(\sin(\theta) - \frac{j\lambda}{d})) \qquad (3)$$

Each ray may intersect a second diffracted ray in a ray tracing extension, giving an intersection point that, being common to most rays, provides a virtual image $A_\lambda$ of the point object. By choosing a pair of rays symmetrically located at a very small angle Δθ on both sides of a central ray incident at the angle θ, we obtain the images at coordinates x, z:

$$x = x_A + z_A \theta + \Delta\theta \left[ \frac{T(\theta + \Delta\theta)\tan(\theta - \Delta\theta) - T(\theta - \Delta\theta)\tan(\theta + \Delta\theta)}{T(\theta + \Delta\theta) - T(\theta - \Delta\theta)} \right] ,$$
$$z = \frac{z_A \tan(\theta) - x + x_A}{T(\theta)} \qquad (4)$$

In this way, we can obtain the position of every diffracted image as a function of the coordinates of the observation point. deriving the proper θ value from the expression

$$\tan(\theta')=\frac{x_F-x_A-z_A\tan(\theta)}{-z_F} \quad , \tag{5}$$

which can be combined with Eq. (1) to give the final expression

$$\arcsin(\sin(\theta)-\frac{j\lambda}{d})=\arctan(\frac{x_F-x_A-z_A\tan(\theta)}{-z_F}) \quad , \tag{6}$$

from which θ can be calculated as a first step for obtaining the x, z values from Eqs. (4).

Table I shows typical calculated values of x,z for the first diffracted order, as a function of θ and Δθ values. Under these conditions, virtual images generated by light impinging at an angle θ=arcsin(λ/d) are seen in a direction that is perpendicular to the plane of the grating, a given distance closer to the observer than the object. These images move closer to or farther from the grating as we look around them. For simplicity, only first order images are considered in this paper, neglecting also the fact that the object may be located outside the x,z plane.

*Table I: Calculated values of the x,z coordinates of the diffracted image as a function of incidence angle θ and beam divergence Δθ. Considered values are d = 1.7 μm. λ= 633 nm, Δθ= 1°, $z_F$ = 255 mm.*

| θ (°) | -x/$z_A$ | z/$z_A$ |
|---|---|---|
| 25 | 0.39 | 0.30 |
| 15 | 0.42 | 0.51 |
| 10 | 0.40 | 0.61 |
| 5 | 0.36 | 0.70 |
| 0 | 0.32 | 0.80 |
| -5 | 0.44 | 0.89 |
| -10 | 0.55 | 0.98 |
| -15 | 0.66 | 1.09 |
| -25 | 0.86 | 1.34 |

Figure 2 shows a clear image obtained in monochromatic light under the experimental conditions that will be described in Sec. 4, within a field of view equal to the angular extension between the extremes of the zero and second order images.

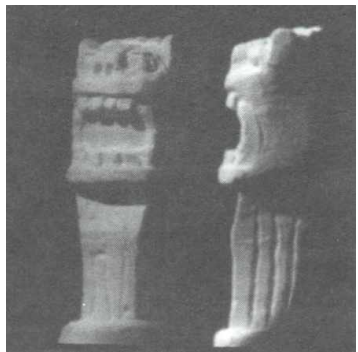

*Fig. 2. Left: Photograph of a nearly cylindrical object of 12 mm diameter, obtained from the first order image of a diffraction grating in white light filtered within a 5 nm bandwidth. Right: Direct (undiffracted) image of the object facing the reflection grating.*

## 3. COMPARISON BETWEEN HOLOPHOTOGRAPHY WITH A DIFFRACTION GRATING

AND CLASSICAL STEREO PHOTOGRAPHY

The wavelength dependence of the image position is the property that allows images of the object to be seen from different points of view. In Fig. 3 we show how θ changes in a two-wavelength case, allowing both wavelengths to reach the same observation point. By considering each ray in the figure as an example of the many rays of the same wavelength that can be collected in that direction, we can understand how the images of any points lying in that line are obtained. The angle of stereopsis is now easily obtained as the difference between the θ angles. Typical values are shown in Table II.

*TABLE II. Valuesof the coordinates of the diffracted imane for three different object points of coordinates $x_A$, $z_A$. Two wavelenths were considered, in a situation that corresponds to the experiment in Fig.4(a).*

|         |         | λ':466nm |     | λ":653nm |      |
|---------|---------|----------|-----|----------|------|
| $x_A$   | $z_A$   | x'       | z'  | x"       | z"   |
| 14      | 51      | -0.7     | 57  | -7.5     | 64   |
| 8.5     | 30.5    | -0.3     | 34  | -4.4     | 38.5 |
| 3       | 10      | 0.1      | 11  | -1.2     | 12   |

The case of holophotography corresponds to the situation of Fig. 3 if we place in position ($x_F$, $z_F$) the entrance pupil of a photographic camera with the image of the reference point O centered on the object field. We consider here, for simplicity, that the coordinates of point O are ($x_F$, 0). The image projected on the photographic film at distance i will register the images $A_r'$ and $A_b'$, of points $A_r$, and $A_b$, separated by the distance

$$\overline{A_r' A_b'} \simeq \frac{i(x_r - x_b)}{z_r - z_F} \quad , \tag{7}$$

where $x_r$, and $z_r$, the coordinates of image point $A_r$, are obtained in the case of red light, while $x_b$, $z_b$, corresponding to the image point $A_b$, are obtained at a shorter wavelength (in blue light, for example).

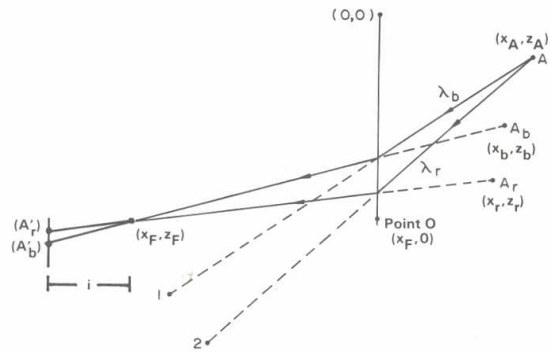

*Fig. 3. Position of the diffracted images according to the wavelength value.*

A double-exposure stereophotographic case is shown schematically in Fig. 1(a) of Ref. 1. The distance between two corresponding points, as registered in the photographs obtained from positions 1 and 2 by centering the camera at the reference point O, corresponds then to the expression

$$\overline{A'A''} = \frac{i d_A \sin\theta_s}{d_A \cos\theta_s - z_F} \quad , \tag{8}$$

where $d_A = [z_A^2 + (x_F - x_A)^2]^{1/2}$ is the distance between points O and A, corresponding to our case, with $\theta_s$ being the angle at which the camera is rotated.

Equations (7) and (8) can now be compared for some typical cases by defining a factor

$$Ms = \frac{\overline{A_r' A_b''}}{\overline{A' A''}} = \frac{x_r - x_b}{z_r - z_F} \frac{d_A \cos\theta_s - z_F}{d_A \sin\theta_s} \quad . \tag{9}$$

By using the values from Table II, $\theta_s = 6.3°$, and $z_F = -530$ mm from Sec. 4, we obtained values $1.15 < Ms < 1.16$ for our experience [Fig. 4(a)]. The difference from the unit value in parameter Ms implies that the angle of stereopsis for observation has changed proportionally, representing a different separation between planes when the object is observed. This result merits further analysis since it seems more appropriate to consider $\theta_s = 7.3°$ for rendering $Ms = -1$.

## 4. EXPERIMENTAL DETAILS

A reflecting ruled grating of 595 lines/mm was used in photographing an object at a distance of 530 mm with a 35 mm camera and a 150 mm focal length objective. The optical axis of the camera was oriented perpendicular to the grating. When the white light was filtered, wavelengths were selected by means of interference filters with a 5 nm bandwidth. The aperture value of the objective was highly reduced in order to yield a good tolerance for the focal depth and corresponded to a $\Delta\theta$ value of approximately $1°$. In Fig.2 the object was centered at the coordinate $z_A = 17$ mm.

Figure 4(a) is a holophotograph of a common steel sewing neddle without its tip, 1.0 mm wide and 62.4 mm long, whose extreme coordinates were (3,10) and (14,51), as expressed in millimeters. The precision of the angle positioning on the rotating table was $0.02°$. The holophotograph is filtered in two waveIengths (466 nm and 553 nm), corresponding to the values in Table II, experimentally verified within a precision of 6%. Some widening in the image is due to reflections on the filters. The red image is at the left and the blue image is at the center. The corresponding nondiffracted image is seen at the right. Figure 4(b) is a double-exposure stereophotograph obtained by performing two simple rotations of the camera position around a point corresponding to reference point O of Fig. 3. The rotation values ($16.1°$ and $22.4°$) were calculated within the scheme of Fig. 3 so as to yield a result very close to that of Fig. 4(a).

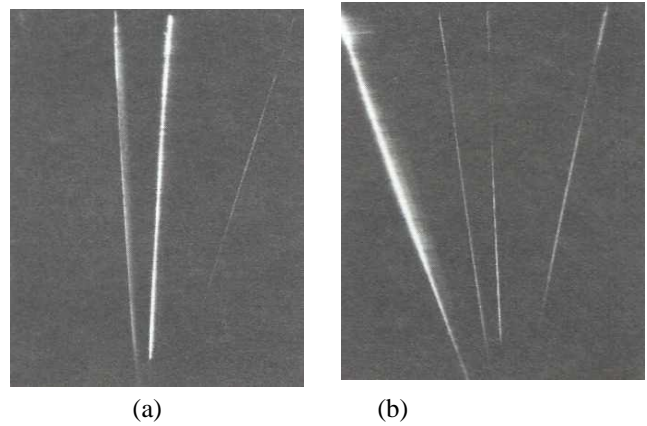

(a)          (b)

*Fig. 4. (a) Photograph of a straight abject (a common sewing needle without its tip) that recedes from the plane of the grating. It was obtained from position ($x_F$, $z_F$), interposing red and blue filters alternately. (b) Double-exposure stereophotograph of the zero order image obtained as in (a) but from positions 1 and 2 in Fig. 3.*

Figure 5 was obtained under conditions similar to those of Fig. 2, but in this case the observation direction is not perpendicular to the grating but at an angle of $24°$. The object was nearly cylindrical, 12 mm in diameter, and also centered at coordinate $z_A = 17$ mm.

*Fig. 5. Holophotography produced using common color film by Illuminating in white light an object that is seen through a diffraction grating. The 3-D effect can be easily observed through colored glasses.*

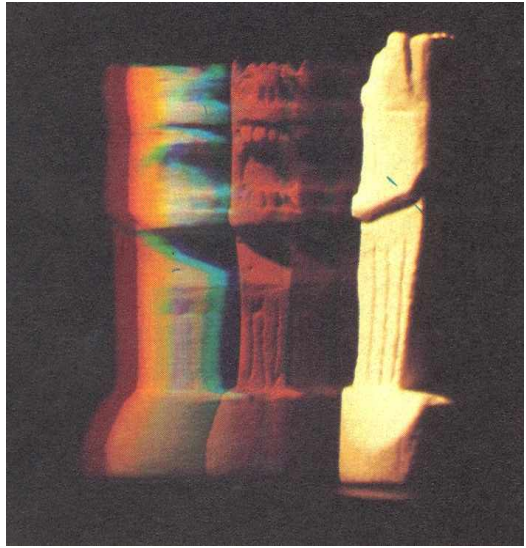

The stereoscopic images were observed in three different ways:

First, the color photographs were observed through absorptive colored filters, yielding a reasonable degree of sharpness and depth perception within a depth of a few centimeters from the plane of the photograph. The depth limitation is due mainly to the limitations of the three-chromatic process of the color film, as described for a similar situation in a previous paper[1].

Second, to observe the images without the limitations of the color response of the photographic film, we projected an image of the first diffracted order onto a diffusing screen, 3 m away from projecting lens of 500 mm focal length at f:8 aperture. By using colored absorption filters, we obtained a great increase in the perception of depth, up to 10 cm.

A further improvement was obtained by observing through 5 nm bandwidth interference filters. This required great visual effort owing to the reduced luminosity of the image. Images could be seen clearly 30 cm in back or in front of the screen; depth reversal was obtained by reversing the filters between the eyes.

In our third observation method, a similar result was obtained by means of a holographic screen that employs the diffraction effect to selectively distribute the images in each direction, according to their color. This will be described in a future paper since it eliminates the need for any glasses, having many important derivations.

In every observation we used common objects, and we noticed no distortion in depth perception, such as a difference in depth other than that normal for those objects. It seems to us that the most fundamental rule in stereoscopic observation is to keep the sequence of the different planes. It is of little effect in the analytical result, indicating that the relative depth between planes could have been modified. From the visual observation, we can say that the 3-D realm of the scene is perfect, as in a conventional hologram.

## 5. PHOTOGRAPHING DIFFRACTED IMAGES: HOLOPHOTOGRAPHY AS THE REGISTERING OF A CONTINUOUS SEQUENCE OF VIEWS

We have seen how a stereophotograph can be obtained with an ordinary photographic camera so as to be seen through ordinary colored 3-D spectacles. Ideally, if the color registering technique were perfect in discriminating and reproducing $\lambda$ values and the color selection by filtering were also perfect, we could register and observe a stereo image, selecting the point of view at will by the choice of the wavelength of the color filter. The situation is equivalent to the one we described previously[1] for the photographs obtained from holograms.

The angle of stereopsis and the object field could be increased by employing gratings with higher spatial frequency. The design of holographic optical elements seems highly indicated in order to improve the possibilities of these elements as diffractive stereo filters.

The images obtained in monochromatic light can be of good quality, but under white light they become blurred because the sensitivity of the color registering process does not discriminate between wavelengths included within the bandwidth of one of the basic colors of the three-chromatic process. The restitution of color after exposure to monochromatic light involves a similar bandwidth. Discrimination of the image by color filters constitutes another imperfect process that makes the resulting photograph interesting in many aspects, as shown in Fig. 5, but not very precise for obtaining clear images in full depth.

Since we have obtained an imaging three-dimensional process that is limited mainly by the color response of the film and filters rather than by geometrical aberrations, it can be improved in many ways. First, a color photographic film could be made with more selective bandwidth discrimination. Not only could conventional color photography be tried, but one could also try the Lippmann process for color reproduction[2], which was previously shown to discriminate at least 10 different bandwidths within the visible spectrum[3]. Second, the observations can be made without any kind of spectacles and with a high degree of color selection by means of a color dispersion process, sending each colored image to a given position for observation. We have developed one process of this kind, which we will report elsewhere.

We predict that the preceding considerations contribute to a photographic technique that may offer three-dimensional images of objects being illuminated in white light in a continuous sequence of views, a feat that previously was possible only by holographic techniques, where the process of interference of light was of basic importance.

## 6. CONCLUSIONS

A photographic process in white light has been shown to share most of the fundamental characteristics of the holographic process of image registering but without requiring a two-beam interference process. The requirements of coherence length for the illuminating source can then be completely neglected. Owing to the perfect capacity of color encoding of views specific to our process, a process of color dispersion can be predicted that, as a complement to this process, may send the images to an observer angularly discriminated according to the original distribution of object views, allowing for a new process of quasi-holographic images being obtained in white light. Formulas were derived to allow calculation of the imaging properties of the registration process.

## 7. ACKNOWLEDGMENT

The author wishes to acknowledge Christiano P. Guerra for his assistance in mounting and photographing the experiments.

January 1990
Volume 29 Number 1
ISSN 0091-3286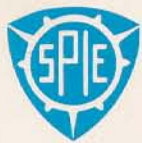

# Optical Engineering

SPIE — The International Society for Optical Engineering

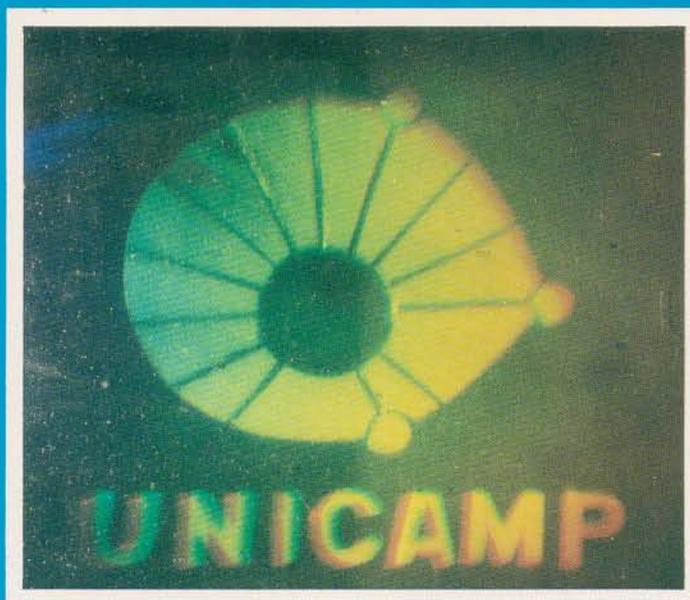

VIEW THROUGH ENCLOSED "3-D" GLASSES (RED—RIGHT EYE)

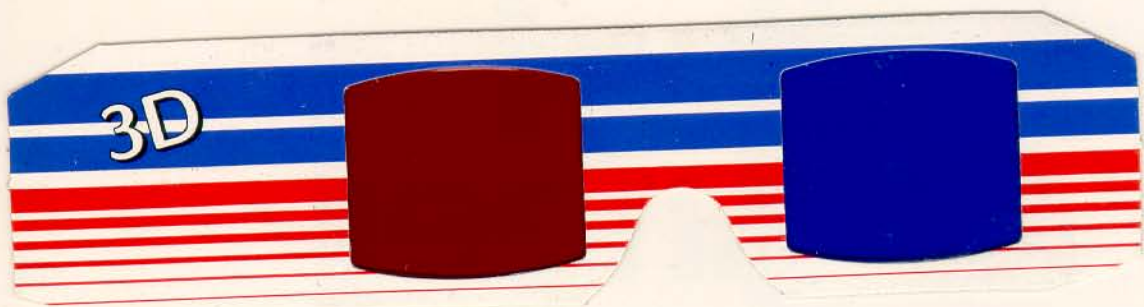

*Editorial*

Jack D. Gaskill, Editor

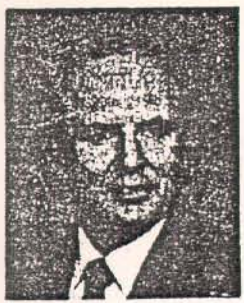

**A Hidden Three-Dimensional Effect**

The photograph on the cover of this issue of *Optical Engineering*, which is taken from Dr. José J. Lunazzi's paper entitled "Three-dimensional photography by holography," contains a hidden three-dimensional effect that can be observed by using the colored spectacles supplied with the issue. This effect is not produced by the usual red-green or blue-green composite photograph, called an anaglyph, but is a natural consequence of the process of recording a color photograph of an image formed by a white-light hologram. Following is a brief account of the events leading up to the publication of José's paper, in which an explanation of the effect is given.

It was over two years ago, while we were standing in the parking lot of the Town and Country Hotel in San Diego, that José introduced the subject of his paper to me. Although I had difficulty understanding the phenomenon he was patiently trying to explain to me, I became quite excited about it. In fact, I found it so fascinating that I immediately sought out two experts in the field of holography to get their opinions regarding the novelty and scientific merit of José's work. When these two experts responded with enthusiasm, I invited José to submit a paper to me for review and possible publication in the Journal, which he did. In addition, he submitted a second, related paper entitled "Holophotography with a diffraction grating," which also appears in this issue.

José first observed this phenomenon in 1984, and he credits his daughter, Silvia, with helping him discover it. He had just returned from an exhibit in Frankfurt, "Licht Blicke," with a catalog that contained over 100 color photographs of images formed by white-light holograms. As he was browsing through the catalog with Silvia, who was nine years old at the time, she suggested that they look at the pictures through some red-green 3-D spectacles that she happened to have. José admits that his first reaction was that of a typical adult—probably something like "oh, good grief"—but that he gave in and did as Silvia wished.

As they started looking at the pictures through the colored spectacles, José noticed that the color filtering produced some "nice" pseudo 3-D effects, but didn't think much about it—yet! Silvia soon tired of this activity, but José pressed on, systematically viewing image after image and occasionally observing an illusion of three-dimensional reality. Then, while viewing the 88th photograph, he suddenly became convinced that this 3-D effect was real—that an authentic representation of depth was present—and that he had discovered a stereo effect in the color photographs of holographic images. This led to several years of research, which yielded the two papers mentioned above.

It has been a lengthy, frustrating process getting these two papers published; José encountered delays in his research, there were postal strikes in Brazil, funding for the color photographs in the text and on the cover was difficult to identify, finding a supplier of the colored spectacles took time, etc. Everything eventually worked out, however, and we finally went to press. We hope that you will find this subject as fascinating and exciting as we did. In conclusion, I would like to thank José for his hard work and patience, Universidade Estadual de Campinas for its support of his research, and the National Science Foundation for providing partial funding for the printing of the color photographs.